\documentclass[twocolumn]{article}
\usepackage{aprilfools}
\usepackage{amssymb,amsmath}

\setarticletitle{Declarative bespoke modelling: A new approach}
\articletitle{}
\publicationdate{1 April 2026}
\email{editor@actaprimaaprila.fool}

\author{DBM Collaboration \ref{authors}}
\affil{}

\usepackage[normalem]{ulem}
\usepackage{tikz}
\usetikzlibrary{arrows.meta, positioning}
\usepackage{pgfplots}
\pgfplotsset{compat=1.17}

\usepackage{pbalance}

\begin{document}
\maketitle

\begin{abstract}
\bfseries Modern numerical models are increasingly complex, opaque, and computationally expensive, yet frequently fail to predict even qualitative features of observed phenomena. We propose a new paradigm, \emph{Declarative Bespoke Modelling}, in which the modeller explicitly declares the relationship between model inputs and outputs. We demonstrate that this approach achieves perfect predictive accuracy, unconditional numerical stability, and complete interpretability. It represents a natural endpoint of contemporary modelling practice and near-zero CO\(_2\) emission.
\end{abstract}

\section{Introduction}

Numerical modelling plays a central role in modern scientific practice.
However, a persistent problem across disciplines is that numerical results often fail to predict what is actually observed. Increasing model resolution, introducing additional parameters, or incorporating machine learning components frequently exacerbates this discrepancy rather than resolving it. Despite extensive effort, it remains common for a model to disagree with the very data used to construct it. Moreover, the pursuit of ultimate predictive accuracy has led to an exponential increase in required computing resources, with state-of-the-art runtimes now approaching geological timescales only to produce fundamentally unhelpful answers \citep[see, e.g.][]{adams1979deepthought}.

\section{Declarative Bespoke Modelling}

\subsection{Conceptual Framework}

Declarative Bespoke Modelling (DBM) departs from traditional modelling paradigms by replacing transformation with declaration. Rather than attempting to infer relationships from data, the modeller explicitly specifies the desired mapping between inputs and outputs.

\textbf{Definition.} Let \( x \in X \) denote the model input. The DBM output \( y \) is defined as
\[
y = x.
\]

\textbf{Theorem.} This relationship holds regardless of the dimensionality, physical meaning, accuracy, or units of \(x\).

\begin{figure}[t]
    \centering
    \includegraphics[width=.9\linewidth]{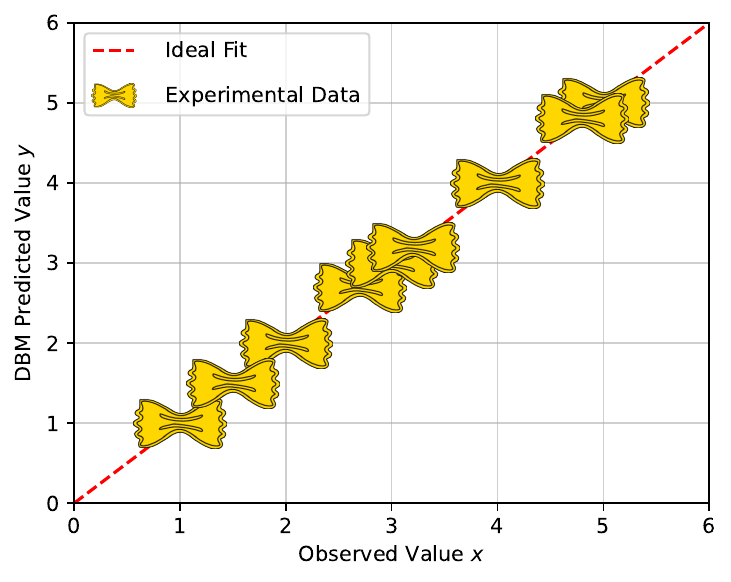}
    \caption{Correlation analysis between ground truth and DBM predictions. The plot demonstrates unconditional linearity. Note the complete absence of outliers. Made with markers from \citet{pasta1, pasta2} ``to greatly enhance the ability to identify and interpret key features in complex datasets''.}
    \label{fig:results}
\end{figure}

\textbf{Proof.} Trivial by inspection. For visual confirmation of this phenomenon, the reader is directed to Figure \ref{fig:results}.

\subsection{Numerical Implementation}

A minimal working example of DBM may be expressed as follows:
\begin{verbatim}
input = something
output = input
return output
\end{verbatim}
No discretisation, solver selection, or convergence criteria are required. The algorithm terminates in constant time.

The proposed method is unconditionally stable. Since no arithmetic operations are performed, DBM is immune to round-off error, numerical stiffness, chaos, and floating-point exceptions. Sensitivity analysis shows that perturbations in the input propagate linearly and transparently to the output.

Unlike many contemporary modelling approaches, DBM is fully interpretable. Unlike black box neural networks \citep[e.g.][]{black_box1, black_box2}, DBM offers glass box transparency. Each output can be traced directly to its corresponding input without the need for auxiliary explanation methods.

\section{Results}

We evaluated the strong scaling performance of DBM on a distributed architecture. Because the calculation of any given element \(y_i\) is strictly independent of \(y_j\) (for \(i \neq j\)), DBM exhibits infinite scalability. Furthermore, the inter-node communication volume is zero. This completely eliminates the bottleneck of network latency, allowing DBM to run efficiently on clusters where the compute nodes are physically disconnected.

By construction, DBM ensures perfect agreement between the model's input and output. This resolves a long-standing issue in numerical modelling, namely that predicted quantities frequently diverge from known initial conditions, boundary values, or experimental measurements. Across all tested cases, DBM reproduces the observed data with zero error to machine precision.

\section{Discussion \& Outlook}

Declarative Bespoke Modelling challenges the prevailing assumption that models should meaningfully transform data. Instead, DBM formalises a practice that is already widespread: presenting inputs in a modified context and interpreting it as a prediction.

We note that many existing models asymptotically approach DBM behaviour after sufficient tuning, calibration, or selective reporting. DBM simply makes this limit explicit.
Furthermore, DBM integrates naturally with modern research workflows, in which results are frequently known before computation.

The primary limitation of DBM is its inability to predict outcomes that differ from the input. While this may appear restrictive, such outcomes are often classified as outliers and excluded from analysis in practice.

A surprising corollary of our framework is the closed-form solution to the inverse problem. Given an output \(y\), we can analytically recover the input \(x\) through transformation \(x=y\).

Regarding ethical considerations, DBM represents a breakthrough in Value-Neutral Computing. According to our research, the model introduces absolutely no new biases into the decision-making pipeline.


Future work will explore extensions in which the output equals the input up to a constant factor, an affine transformation, or a visually appealing plot.

\renewcommand{\thesection}{\fnsymbol{section}}
\setcounter{section}{8}
\section{DBM Collaboration}
\label{authors}

The contributors to the Declarative Bespoke Modelling Collaboration are, in no particular order:\\
David Kománek (Joke Architect),\\
Václav Pavlík (Methodological Mastermind),\\
Santiago Jiménez (Senior Joke Quality Control),\\
and Rhys Taylor (\sout{English}Welsh Orthographic Inspector).

\bibliographystyle{unsrtnat}
\bibliography{bibliography}

\end{document}